\documentstyle[jkas]{article}

\beginpage{1}
\endpage{4}
\year{2011}\volume{43}\month{February}

\runningauthor {B. LIM ET AL.} 
\runningtitle{THE OPEN CLUSTER NGC 2353}

\month{1} \year{2011} \volume{1}
\beginpage{1}\endpage{2}
\date{2011}

\begin{document}
\title{SEJONG OPEN CLUSTER SURVEY. I. NGC 2353}
\author{Beomdu Lim$^{1}$, Hwankyung Sung$^1$, R. Karimov$^2$, M. Ibrahimov$^2$}
\address{$^1$ Department of Astronomy and Space Science, Sejong University, Seoul, Korea\\
 {\it E-mail : bdlim1210@sju.ac.kr, sungh@sejong.ac.kr}}
\address{$^2$ Ulugh Beg Astronomical Institute, 33 Astronomical Street, Tashkent 700052, Uzbekistan\\
{\it E-mail : rivkat@astrin.uzsci.net, mansur@astrin.uzsci.net}}

\address{\normalsize{\it (Received ??, ??; Revised ??, 2011; Accepted ??, 2011)}}
\offprints{H. Sung}
\abstract{$UBVI$ CCD photometry of NGC 2353 is obtained as part of the ``Sejong Open cluster
Survey" (SOS). Using the photometric membership criteria we selelct probable members of the cluster.
We derive the reddening and distance to the cluster, i.e. $E(B-V) = 0.10 \pm 0.02$ mag and $1.17 \pm 0.04$
kpc, respectively. We find that the projected distribution of the probable
members on the sky is elliptical in shape rather than circular. The age of the cluster is estimated to be
log(age)=8.1 $\pm$ 0.1, older than what was found in previous studies. The minimum value of binary
fraction is estimated to be about 48 $\pm$ 5 percent from a Gaussian function fit to the distribution
of the distance moduli of the photometric members. Finally, we also obtain the luminosity function and the
initial mass function (IMF). The slope of the IMF is $\Gamma = -1.3 \pm 0.2$.}

\maketitle

\section{INTRODUCTION}
NGC 2353, known as a young open cluster, is located in the
easthern edge of the Canis Majoris OB1 association (hereafter CMa OB1). In spite
of its proximity only a few studies were devoted to the cluster, because most of the research 
up to date focused on the star-forming regions in the CMa OB1.

\citet{H61} observed 70 galactic clusters in the $UBV$ system. NGC 2353
is one of their targets. \citet{H65} derived the distance moduli of open clusters using the
zero-age main sequence (ZAMS) relation, the photometric $H\gamma$ equivalent width, and
spectral classifications. They estimated the distance modulus of NGC 2353 to be 10.4 mag. \citet{FHR90} (hereafter FHR90) studied NGC 2353 using their
$UBV$ photoelectric  photometry of 46 stars. They derived the distance modulus ($V_0 - M_V = 10.39 \pm 0.08$), 
the foreground reddening ($<E(B-V)> = 0.12 \pm 0.04$), and the binary fraction ($33 \%$) of the cluster.

An interesting aspect of NGC 2353 is whether or not it is physically connected with CMa OB1. 
\citet{A49} firstly suggested this hypothesis. Subsequently \citet{R66} proposed that NGC 2353 may be the nucleus of the CMa
OB1, because of their similar distance ($d$ = 1.3 kpc; \citealt{B63}). Later, \citet{C74} compared
the age of CMa OB1 (3 Myr) with that of NGC 2353 (12.6 Myr -- \citealt{H61}) and suggested that
the first star formation in the region was ignited at NGC 2353, and then
propagated to the present center of CMa OB1. On the other hand, FHR90 argued that
they could not find any physical relationship between NGC 2353 and CMa OB1 from the derived
age of the cluster (76 Myr). Finally, \citet{E78} suggested that NGC 2353 may not be a physical group, 
by analyzing the $uvby$ and H$\beta$ photometry of each group.

We are running an open cluster survey program, i.e. the ``Sejong Open
cluster Survey" (SOS). The main aim of the project is to provide homogeneous photometric
data (see \citealt{MP03}) in the Johnson-Cousins' $UBVI$ system, which is tightly matched to
the SAAO standard system (\citealt{M89},1991; \citealt{K98}; see also \citealt{L09}). The 
collected homogeneous data will be used to revise and update the spiral arm structure of the Galaxy as well
as the star formation history in the Galaxy. The data can also be used to test observationally the stellar
evolution theory. The project overview and calibrations for data analysis purposes will be presented 
elsewhere \citep{L11}. In this paper we describe the first results from the SOS.

NGC 2353 is chosen as a target of the SOS because the cluster is relatively close and rich. However, 
no CCD photometry has been performed so far. In addition, there is a problem related with its age 
-- the age of NGC 2353 in the open cluster data base WEBDA\footnote{http://www.univie.ac.at/webda/} is about
$10^8$ yrs, but an evolved O type star HD 55879 (Sp: O9.5 II-III) is located near the
center of the cluster. The relationship between NGC 2353 and CMa OB1 is also still an open issue, to be 
clarified from this study. In Section 2, we present the observations and comparison with previous photometry. 
We provide $UBVI$ CCD photometry for the stars in the cluster up to $V=20$ mag. In Section 3, we present 
photometric diagrams and several fundamental parameters derived from  them. Luminosity and mass functions are presented in
Section 4. Finally, we discuss the binary fraction and background population in Section 5. We highlight our conclusions
are in Section 6.

\begin{figure}[!t]
\centering
\includegraphics[width=9cm]{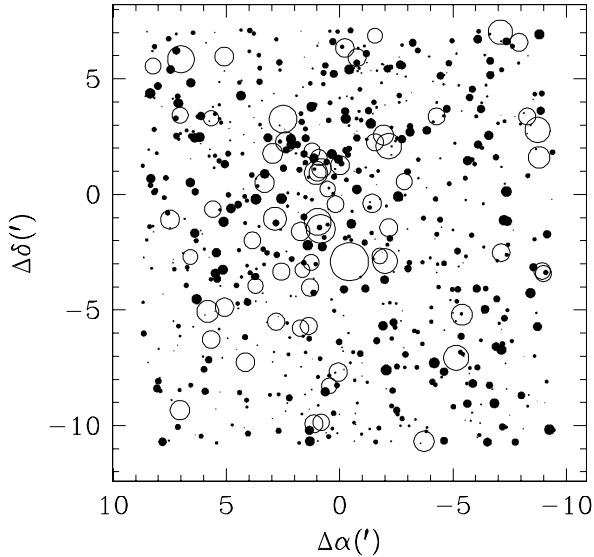}
\caption{Finding chart for NGC 2353. The size of the circle is proportional
to the brightness of the star. The position of the stars is relative to the
center of the cluster ($\alpha = 7^h 14^m 30^s.0$ $\delta = -10^{\circ}16.1$, J2000). \label{fig1}}
\label{fig1}
\end{figure}

\section{OBSERVATIONS}
The observation of NGC 2353 was made on 2007 January 25 with the AZT-22 1.5m telescope
(f/7.74) at Maidanak Astronomical Observatory in Uzbekistan. All the images were
taken with a Fairchild $4098 \times 4098$ CCD \citep{I10} and standard Bessell $UBVI$ filters \citep{Be90}.
The pixel scale is $0.{\arcsec}266$ $pix^{-1}$, and therefore a side of the CCD
corresponds to $18\arcmin.1$. Two sets of exposure times for each band for a total of 8 frames were taken
in the observation -- 60s and 3s in $I$, 180s and 5s in $V$, 300s and 7s in $B$, 600s and 15s in $U$.
We present the finding chart for the stars brighter than $V=17$ mag in Fig. \ref{fig1}.
Photometric data for five stars saturated in our images are taken from WEBDA, and averaged.

The removal of instrumental artifacts is performed with the IRAF/CCDRED package, 
as described in \citet{L08}. PSF photometry was performed for all the stars, using IRAF/DAOPHOT.
Many equatorial standard stars in \citet{M91} were observed on the same night in a wide
range of air masses to derive the atmospheric extinction coefficients and the photometric
zero points. We present the coefficients and the photometric zero points in Table \ref{tab2}.
The instrumental magnitudes were converted into the standard system following the procedure
described in \citet{L09}.

\begin{table}[t]
\begin{center}
\scriptsize
\caption{ Photometric data for five bright stars from WEBDA \label{tab1}}
\renewcommand\arraystretch{1.5}
\begin{tabular}{ccccc}
\hline \hline
ID & $V$ & $B-V$ & $U-B$ & $n_{pe}$\\
\hline
HD 55879 & 6.012 & -0.171 & -0.991 & 7 \\
HD 55930 & 9.159 & -0.014 & -0.291 & 3 \\
HD 56009 & 9.450 &  0.977 & 0.698  & 3 \\
HD 55901 & 8.826 & -0.036 & -0.530 & 3 \\
BD-10 1935 & 9.380 &  0.005 & -0.330 & 2 \\
\hline\\
\end{tabular}
\end{center}
\end{table}

\begin{table}[!t]
\begin{center}
\scriptsize
\caption{ Extinction coefficients and photometric zeropoints\label{tab2}}
\renewcommand\arraystretch{1.5}
\begin{tabular}{ccccc}
\hline \hline
Filter & Color & $k_{1\lambda}$ & $k_{2\lambda}$ & $\zeta_{\lambda} (+25 mag)$\\
\hline
$I$ & $V-I$ & $0.041 \pm 0.013$ &         -         & $-1.772 \pm 0.015$  \\
$V$ & $V-I$ & $0.111 \pm 0.008$ &         -         & $-1.360 \pm 0.010$  \\
$V$ & $B-V$ & $0.111 \pm 0.008$ &         -         & $-1.364 \pm 0.010$  \\
$B$ & $B-V$ & $0.231 \pm 0.009$ & $0.034 \pm 0.003$ & $-1.519 \pm 0.012$  \\
$U$ & $U-B$ & $0.405 \pm 0.009$ & $0.016 \pm 0.003$ & $-3.309 \pm 0.016$  \\
\hline\\
\end{tabular}
\end{center}
\end{table}

\begin{figure*}[!t]
\centering
\includegraphics[width=13cm]{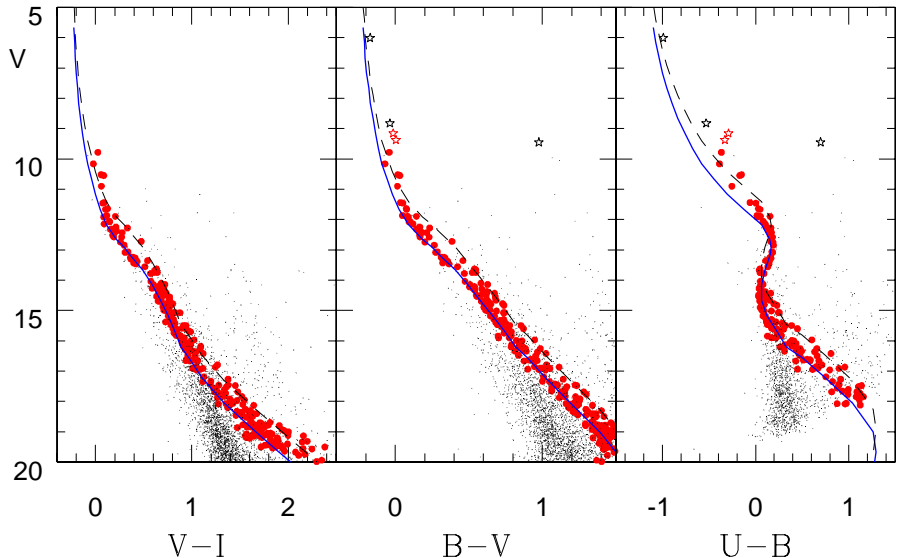}
\caption{Color-magnitude diagrams of NGC 2353. The solid (blue) line represents the reddened zero-age main sequence relation \citep{SB99} where 
$E(B-V) = 0.10$ and $V_0 - M_V = 10.35$. The dashed line denotes that for equal mass binaries. A 
large dot (including a red star mark) or a star mark denotes a probable member of NGC 2353, respectively, 
based on photometric criteria and photometric data obtained from previous photoelectric photometry.
\label{fig2}}
\label{fig-single}
\end{figure*}

\begin{table*}[!!t]
\begin{center}
\caption{Comparisons with previous photometry\label{tab3}}
\renewcommand\arraystretch{1.5}
\begin{tabular}{ccccccc}
\hline \hline
Reference & $\Delta V$ & $n(n_{ex})$ &$\Delta (B-V)$ & $n(n_{ex})$ &$\Delta (U-B)$ & $n(n_{ex})$ \\
\hline
\citealt{H61} & $ 0.010 \pm 0.047$ & $ 5(0)$ & $-0.016 \pm 0.037$ & $ 5(0)$ & $ 0.031 \pm 0.084$ & $ 5(0)$ \\
\citealt{FHR90} & $-0.003 \pm 0.038$ & $10(2)$ & $ 0.013 \pm 0.051$ & $11(1)$ & $-0.025 \pm 0.057$ & $11(1)$ \\
\hline\\
\end{tabular}
\end{center}
\end{table*}

In order to check the consistency of our photometry with results from previous studies, 
we present the differences and the standard deviation in Table \ref{tab3}. Since the published
photoelectric data contain mostly bright stars, only a small number of
stars are used in the calculation. We use the symbol $\Delta$ to denote the difference between previous 
photometry and our data; $n$ represents the number of stars included in the computation, 
while $n_{ex}$ denotes the number of stars excluded above the $2\sigma$ level. Our photometry shows 
good agreement with previous photometry in $V$ and $B-V$. Although there is a slight systematic 
difference of $0.02$ mag between FHR90 and our photometry in $U-B$, the difference is not signigicant. 
It is hard to compare with \citet{H61}, because of the small number of stars in common. However, the 
difference is also acceptable since the standard deviation indicates that the dispersion is not 
systematic, and our $U-B$ data follow rather well the intrinsic color relation (see Fig. \ref{fig3} or \citealt{SB99}, hereafter SB99).
Photometric data are available at http://arcsec.sejong.ac.kr/$\sim$ sungh/paper.html.

\section{PHOTOMETRIC DIAGRAMS AND FUNDAMENTAL PARAMETERS}
We present three color-magnitude diagrams (CMDs) in Fig. \ref{fig2}. The well-defined
main sequence can be clearly recognized, although a number of background stars in the 
line of sight are superimposed. Some of them are main sequence (MS) turn-off stars in 
the Perseus arm, passing through the MS band of NGC 2353. Several foreground stars, and 
probable members of CMa OB1, seem to populate the brighter part of the CMD.

\begin{figure}[!!h]
\centering
\includegraphics[width=7.5cm]{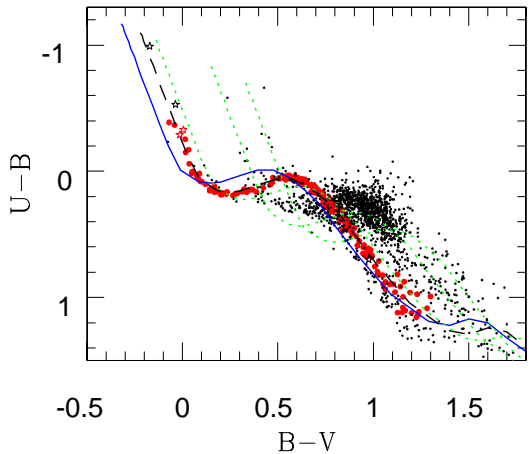}
\caption{Color-color diagram of NGC 2353. The solid line (blue), the
dashed line, and the dotted line (green) represent respectively the intrinsic 
color-color relation \citep{SB99}, the reddened color-color relation for NGC 2353, 
and the reddened color-color relations for three background groups. Symbols have 
the same meaning as in Fig. \ref{fig2}. \label{fig3}}
\label{fig-single}
\end{figure}

In the color-color diagram of Fig. \ref{fig3}, there are 4 groups of early-type stars 
with different reddening. This fact indicates that several young populations
exist in the line of sight. FHR90 already reported two groups, i.e. Group A and B. According 
to their study, stars which belong to Group A are more reddened than those of NGC 2353 
and may be related to CMa OB1. On the opposite, highly reddened Group B stars may be early-type stars distributed along
the line of sight to the outer disks of the Galaxy. We will discuss
these background populations in detail in Section 5.

\subsection{REDDENING LAW}
In general, the interstellar reddening is determined by comparing the observed $B-V$ and
$U-B$ colors with the intrinsic color relation of early-type stars. We use those of SB99 
in the reddening determination. The reddening slope is assumed to be $E(U-B)/E(B-V) = 0.72$. 
In order to select early-type members used for determining of the reddening, we adopt the following 
criteria; (1) $B-V \le 0.09$, (2) $U-B \le 0.06$, (3) $V \ge 9.0$ mag. We estimate the mean interstellar 
reddening to be $E(B-V) = 0.10 \pm 0.02$ (s.d.), using 13 late B type members selected with the 
previous criteria. Our result is in good agreement with \citet{J61} $[E(B-V) = 0.12]$ and FHR90 
(i.e. within the standard deviation). If we adopt the ratio of the total to selective extinction to be 
$R_V = 3.1$, the total extinction in $V$ is about $0.31$ mag.

\begin{figure}[!!t]
\centering
\includegraphics[width=8cm]{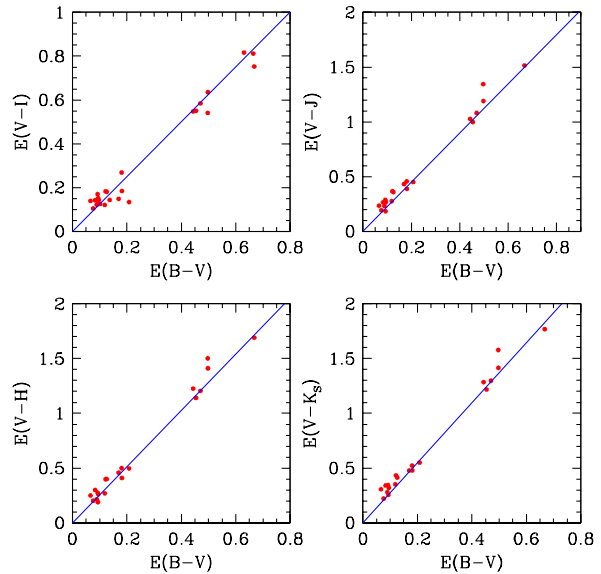}
\caption{Reddening law of the NGC 2353 region. The (red) dots and the solid line represent 
respectively the color excess ratio of early type stars in the observed region, and the 
standard reddening law between two color indices. \label{fig4}}
\label{fig-single}
\end{figure}

In order to check the reddening law, we plot the color excess ratios in Fig. \ref{fig4} using both our
photometric data and the 2MASS near-IR data \citep{C06}. The normal values of the color excess ratio from \citet{GV89}
are $E(V-I)/E(B-V)=1.25$, $E(V-J)/E(B-V)=2.24$, $E(V-H)/E(B-V)=2.57$, $E(V-K)/E(B-V)=2.74$. We transform 
CCD coordinates ($x_{CCD}, y_{CCD}$) into the celestial coordinates ($\Delta \alpha$, $\Delta \delta$)
for the stars matched with the object in the 2MASS catalogue within $1\arcsec$. The symbol $\Delta$ is 
used to indicate the relative position from the adopted cluster center. The early type stars in the observed region show
a wide range of reddening. All groups follow the standard reddening law in $E(V-I)/E(B-V)$ (upper left panel). 
Hence, our photometry is well tied to the standard system and therefore the
values of $E(B-V)$ are well determined. The color excess ratio of $E(V-H)/E(B-V)$ (lower left
panel) also shows a good agreement with the standard relation. The small systematic deviation
in $E(V-J)/E(B-V)$ and $E(V-K)/E(B-V)$ might be caused by an error of about 0.03 mag in the
photometric zero points in the 2MASS $J$ and $K$ bands. From Fig. \ref{fig4} we conclude that the 
reddening law in the direction of NGC 2353 is normal ($R_V = 3.1$).

\subsection{MEMBERSHIP SELECTION AND DISTANCE MODULUS}
Membership selection is crucial in the study of open clusters. Most of the open clusters are 
located in the Galactic plane, therefore we expect significant foreground and background field 
contamination. Proper motion study is essential for membership selection, and provides a 
valuable independent information. Unfortunately membership selection from proper motion study 
is inaccurate, because many photographic plates taken at different epochs spread over a few 
decades or up to a century are required to achieve good accuracy. In addition, these photographic plates
are very rare and the photometric depth is very shallow due to low sensitivity and small 
dynamic range of the old photographic plates.

\begin{figure}[!!t]
\centering
\includegraphics[width=8cm]{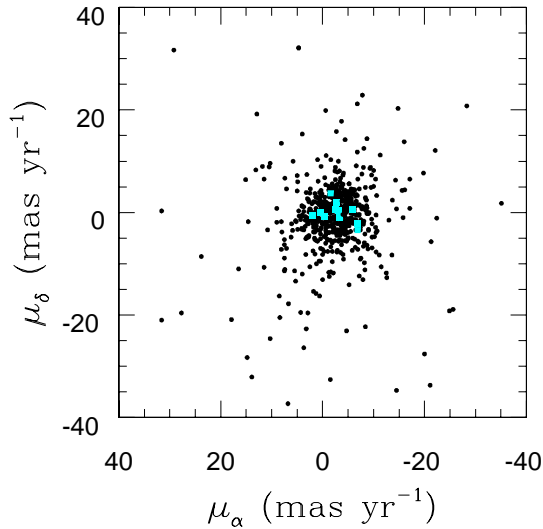}
\caption{Distribution of the proper motions of stars within $1\arcdeg$ from the
center of NGC 2353. Cyan (lightly shaded) dots denote the photometric members of NGC 2353.
Most of the stars around NGC 2353 have similar proper motion. \label{fig5}}
\label{fig-single}
\end{figure}

We select proper motion members of NGC 2353 using proper motion data from the Tycho 2 catalogue 
\citep{T00}. The proper motion of stars in the field of NGC 2353 is shown in Fig. \ref{fig5}. 
Unfortunately, the mean proper motion of the suspected members is very
similar to that of the field stars at almost the same distance (members of the CMa OB1 association). 
Therefore, it is not feasible to select proper motion members from the Tycho catalogue. If we used 
the proper motion data from UCAC3 \citep{Z10}, the result would be very similar.

SB99 devised photometric membership criteria which use multicolor photometry (see also \citealt{K10}). 
The distance moduli of the stars in a cluster derived from different color indices should be consistent. 
However the difference in abundance (SB99) or in the level of chromospheric activity (\citealt{SB02} -- SBLL02) 
makes the MS band bluer, especially in the ($M_V,B-V$) relation. For these cases, we should modify the
ZAMS relation before applying the membership criteria.

\begin{figure}[!]
\centering
\includegraphics[width=8cm]{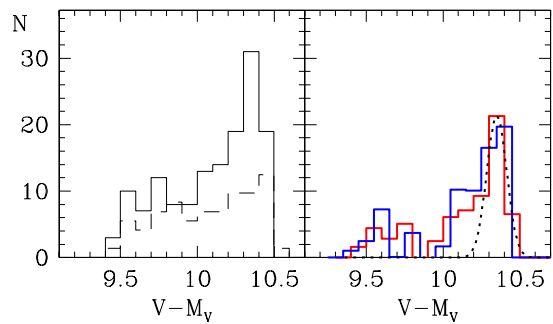}
\caption{Distribution of distance moduli. The thin solid and dashed histograms represent
the number of photometric members within NGC 2353 and the number of stars outside the cluster,
respectively. The thick solid histograms (red and blue) show the distribution of distance moduli
after subtraction of the contribution of field stars. Only stars with $V = 13 < V < 18$ mag are considered
to avoid the effects of evolution. A Gaussian function fit to the last three bins
($10.3 \le V_0 - M_V \le 10.5$) provides the distance modulus of NGC 2353. \label{fig6}}
\label{fig-single}
\end{figure}

Luckily, for the case of NGC 2353 we do not find signatures of non-solar abundance. We calculate
the average value and the difference of the distance moduli for individual stars. We select members
of NGC 2353 if (1) the average value of the distance modulus is between $[(V_0-M_V)_{cluster}
- 0.75 - 2\sigma _{V_0-M_V}]$ and $[(V_0-M_V)_{cluster} + 2\sigma _{V_0-M_V}]$, and (2) if the 
difference in distance moduli is less than $\pm 2.5\sigma _{V_0-M_V}$ where $\sigma _{V_0-M_V}$ is
the width of the Gaussian fit. We plot the distance modulus distribution of the photometric
members in Fig. \ref{fig6}. The bin size is 0.1 mag. Stars with $V =$ 13 -- 18 are used in the
histrogram to avoid the effects of evolution.

We statistically subtract the expected number of field stars within the cluster area which meet the
photometric membership criteria (see the left panel of Fig. \ref{fig6}). The residual histogram
(red) is plotted in the right panel. We also superimposed an histogram (blue) shifted by 0.5 mag to minimize
binning effects. A distance modulus of $10.35 \pm 0.08$ mag is derived from a Gaussian function
fit to the distribution. This value is in good agreement with the distance modulus determined by
FHR90 or \cite{H61}.

The criteria (1) and (2) are not applied to several evolved stars ($V\le13$). Although
the photometric membership criteria constrain well the members, many field stars may also be
included if the differences in reddening and distance are very similar to those of the cluster
members. In order to eliminate these interlopers, we use another criterion - the
($U-B, B-V$) diagram in Fig. \ref{fig3}. We only select the stars within $\pm 2.5\sigma_{CC}$
from the reddened color-color relation shown in Fig. \ref{fig3}. The quantity $\sigma_{CC}$
= $\sqrt{\sigma^2_{B-V} + \sigma^2_{U-B}+\sigma^2_{E(B-V)}}$ represents the total error in the calculation.

\begin{figure}[!!!t]
\centering
\includegraphics[width=8cm]{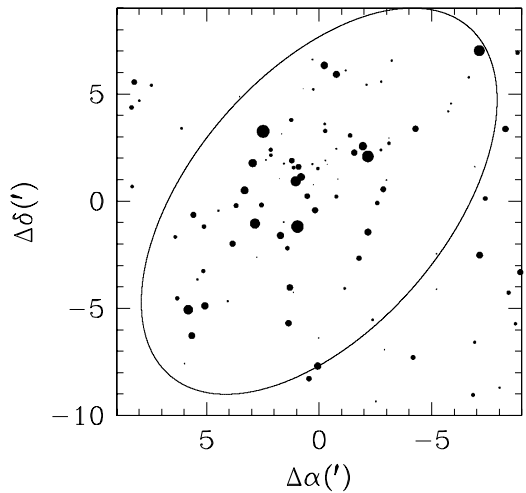}
\caption{Spatial distribution (left) and surface density (right) of the bright stars selected using 
the photometric membership criteria. The size of the dot is proportional to the brightness of the star. 
The shape of the distribution is elongated along the axis connecting from northwest
to southeast. This feature can also be seen in the contour plot. An ellipse
(semi-major axis $= 10\arcmin.5$, eccentricity $= 0.83$, and rotating angle = $52^{\circ}$)
is considered to be the spatial extention of NGC 2353.\label{fig7}}\label{fig-single}
\end{figure}

FHR90 counted the number of stars within a given ring by assuming NGC 2353 to have circular shape, 
and then determined the diameter of the core ($d_{core}=5\arcmin$ ) and of the halo ($d_{halo}=10\arcmin$).
They confirmed that the number of members derived from the photometric criteria and that from the statistical
method is consistent. Fig. \ref{fig7} shows the distribution of bright photometric members
($V \le 15$ mag). Clearly their distribution is elongated like an ellipse. The elliptical
shape can be used to constrain the member distribution of NGC 2353. Several stars
outside the ellipse also meet the photometric membership criteria. However, it is very difficult
to discriminate between members and field stars from the photometry alone, if the photometric characteristics
of the field stars are very similar to those of the cluster stars.

\begin{figure*}[!!!t]
\centering
\includegraphics[width=13cm]{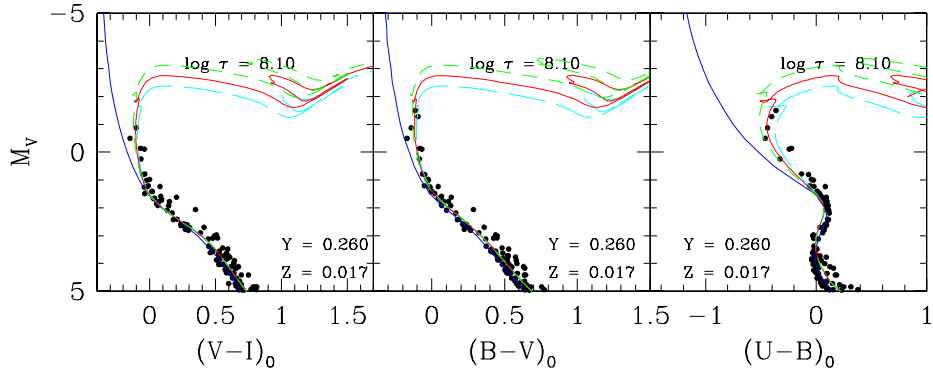}
\caption{Age determination for NGC 2353. Several isochrones [log$t=8.0$ (green), log$t=8.1$ (red), and
log$t=8.2$ (cyan) with $Y=0.26$, and $Z=0.17$] from the Padua group are superimposed to the CMDs.
The solid line (blue) below these isochrones by about 0.3 mag represents the ZAMS
relation of SB99. \label{fig8}}
\label{fig-single}
\end{figure*}

Three bright stars (HD 55879, 56009 and 55901) are excluded from the member list. Since
HD 55879 (its spectral type is O9.5II-III, -- \citealt{W72}) is about 4 magnitude brighter
in $V$ than the MS turn-off ($V_{MSTO} \approx 10.2$ mag) of NGC 2353, the star may be a member of CMa OB1 association.
The star HD 55901 is not only slightly brighter than the MS turn-off, but also bluer
in $U-B$. Although the star may be a blue straggler, we cannot exclude the possibility
that it may be a member of the CMa OB1 association. A giant star HD 56009 is slightly
fainter than the brightness predicted from the Padua theoretical stellar evolution models 
in the observational H-R diagrams in Fig. \ref{fig8}. The membership of HD 56009 and HD 55901 should be checked 
by other independent methods, such as radial velocity measurements.

\begin{figure*}[!!t]
\centering
\includegraphics[width=12.5cm]{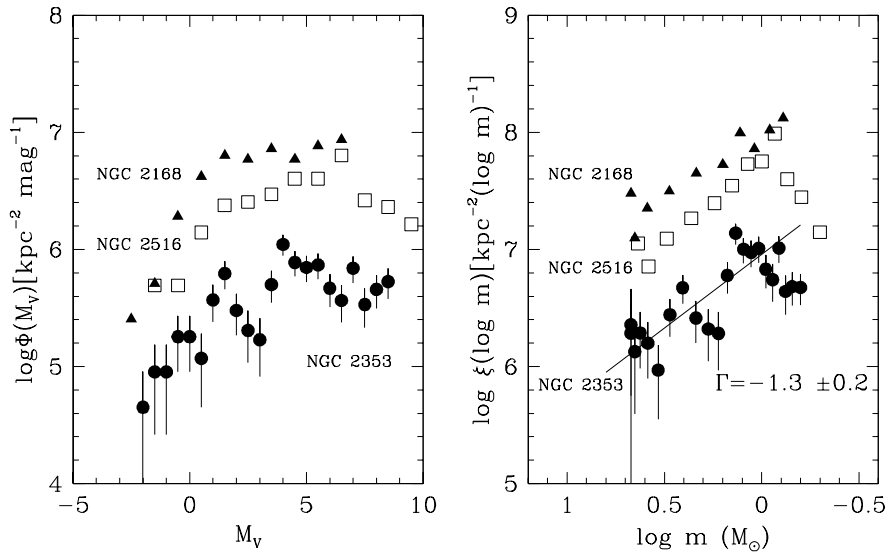}
\caption{Luminosity function (left) and mass function (right) of NGC 2353. The large dot
denotes the LF and MF of NGC 2353. See the main text for details on the LF. The LF and MF 
of NGC 2168 and NGC 2516 are also plotted, for comparison.
\label{fig9}}
\label{fig-single}
\end{figure*}

\subsection{AGE}
The age of a stellar system is generally estimated from the comparison between the
observed CMDs and the theoretical isochrones. Fig. \ref{fig8} shows the observational
H-R diagrams of NGC 2353. Solid lines are isochrones from the Padua group (\citealt{Be08}, 2009)
with solar abundance ($Y=0.26,Z=0.017$) and the convective overshooting parameter
$\Lambda_{C} =$ 0.5. Despite a few giants are found in the observed field, those are 
too faint to be members of NGC 2353. The absence of giants makes it
difficult to constrain the age accurately in the ($M_V,(B-V)_0$), or in the ($M_V,(V-I)_0$) plane.
Fortunately, unlike NGC 2516 (SBLL02) and NGC 2168 (SB99) the MS turn-off
of NGC 2353 is clearly seen in ($M_V,(U-B)_0$). The isochrone of age log $\tau = 8.1 \pm 0.1$ fits
well to the observed sequence in NGC 2353. Although NGC 2353 is slightly younger than
NGC 2516, NGC 2353, NGC 2168 and NGC 2516 can be treated as the same age group.

A physical connection between NGC 2353 and the CMa OB1 has been debated by several authors, as
mentioned in Section 1. FHR90 provided a clue for the large difference in age between
both groups, and suggested that NGC 2353 is not originated from the same cloud. According to
\citet{LBT89}, the natal cloud begins to undergo gas depletion when the age of the cluster
is about 10 Myr. The hypothesis that NGC 2353 is associated with the cloud surrounding the
CMa OB1 (\citealt{A49};\citealt{C74}; \citealt{R66}) cannot explain the large difference
in age. In this respect, our result supports the claim of FHR90.

\section{LUMINOSITY AND MASS FUNCTIONS}
We derive the luminosity function (LF) of NGC 2353 using the photometric members. The fainter
limit in the LF strongly depends on the completeness in a given range of
magnitude. In the region around NGC 2353 crowding effect is not too severe to miss
a number of faint stars. Although several faint stars are affected by the wing
or the spike of bright stars, the number of those stars is not significant. In order
to estimate the completeness limit of our photometry, we check the turnover points
in the distribution of magnitude for a given filter. The completeness limit in
$B,V$ and $I$ is 19.5 mag, 18.5 mag, and 17.5 mag, respectively. Our photometry
is nearly complete down to 18.5 mag in $V$, or equivalently 7.8 mag in $M_V$.

The number of stars in each magnitude bin is counted, and then divided by the area
of NGC 2353. The bin size $\Delta M_V$ is 1 mag. We compare 
the LF of NGC 2353 with that of NGC 2168 and NGC 2516 in Fig. \ref{fig9}. Since the
LFs of NGC 2168 and NGC 2516 are also normalized by their area, the LF represents
the surface number density, i.e. Fig. \ref{fig9} represents the surface density of the member stars
for a given magnitude interval. The number of stars in NGC 2168 per unit area is the highest
in the whole range of magnitude, while that of NGC 2353 is the lowest. This is simply
due to the spatial coverage of the observed region. For NGC 2168, SB99 observed only
the central $20.\arcmin5 \times 20.\arcmin 5$ of the cluster (diameter $\approx 1\arcdeg$).
For NGC 2353, our observed region covers nearly the whole area of the cluster. There is a dip
near $M_V = 3$ in the LF of NGC 2353 (i.e. there are two bumps
in the LF at $M_V$ = 1.5 and at 4). The same trend can also be seen in Fig. \ref{fig8}, 
for stars near $(B-V)_0=0.3$ or $(V-I)_0=0.4$, and $M_V=3.2$, which
corresponds to late A or early F stars, are obviously underpopulated. FHR90 found
a similar gap at slightly fainter absolute magnitude, and suggested that the paucity of A-F type
stars may be related to the spectral peculiarities found from spectra of several
stars in the cluster.

The LF of NGC 2353 can be transformed into the mass function (MF) using the relation
between LF and MF, i.e. $\xi$(log $m$) dlog $m$ $\equiv$ $\Phi(M_V)$ d$M_V$. In the calculation
of MF, the mass-luminosity relation from the appropriate isochrone (log$\tau=8.1$ -- \citealt{Be08}, 2009) is
required. In deriving the slope of the MF, the last two bins are excluded because of
incompleteness of the photometry. The slope $\Gamma =-1.3 \pm 0.2$ is determined with 
the linear least square method. This value is in good agreement with that of NGC 2516 (
$\Gamma = -1.4 \pm 0.3$ -- SBLL02), the solar neighborhood ($\Gamma = -1.35$ -- \citealt{S55}),
and Pleiades and Praecepe ($\Gamma = -1.5 \pm 0.3$ -- \citealt{RM98a}, 1998b), but
it is slightly steeper than that of NGC 2168 ($\Gamma = -1.1 \pm 0.3$ -- SB99). \citet{S99}
found a steep slope for the MF of M11 ($\Gamma = -2.0 \pm 0.6$). The slopes of MF
for the same age group seem to be similar, within errors. However, there may be differences 
between the central outer regions due to the effect of mass segregation
(\citealt{S99}; \citealt{RM98a},1998b).

\begin{figure*}[!t]
\centering
\includegraphics[width=11cm]{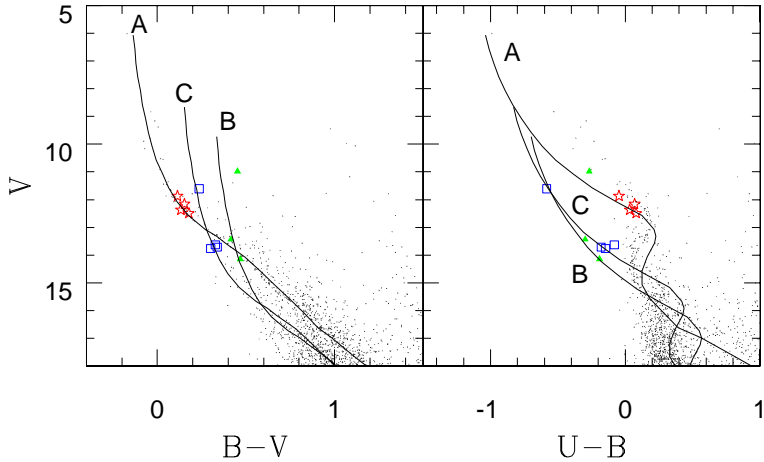}
\caption{Distance moduli of the background groups. The solid line represents the reddened ZAMS
relation, shifted by reddening and distance of each group. The star mark (red), the filled
triangle (green), and the open square (blue), denote Group A, B, and C, respectively. The
Groups A and B correspond to those of FHR90.
\label{fig10}}
\label{fig-single}
\end{figure*}

\section{Discussion}
\subsection{Binary fraction}
The binary fraction and the mass ratio distribution among stars in a cluster 
are crucial for understanding the star formation processes. In addition,
binaries play an important role in the dynamical evolution of a cluster. In order
to estimate the minimum value of binary fraction in NGC 2353, we follow the
same procedure as in SB99. Among 66.4 statistically corrected cluster stars
($V =$ 13--18 mag) in Fig. \ref{fig6}, the number of stars encompassed by the 
Gaussian fit is 34.5. The number of binary
stars which cannot be considered for a Gaussian function fit is thus
31.9. The error of binary fraction can be estimated from the standard deviation
of the residuals in the last four bins. We found a minimum binary fraction
of 48 $\pm 5 \%$ in NGC 2353. The binary fraction obtained is larger
than that of FHR90 (33\%) for the cluster. SB99 found a minimum binary fraction
of 35 $\pm 5\%$ for NGC 2168, and SBLL02 found $40 \pm 4\%$ for NGC 2516. Our result
for NGC 2353 is larger than that of other clusters, while it is rather similar
to the values of 48\% and 43\% in the center of the Pleiades and Praecepe (\citealt{RM98a},1998b)

According to \citet{Mer92} and \citet{Pe98}, a number of spectroscopic binaries
can exist near the ZAMS without a brightening effect, and therefore
our estimate may be a minimum value. Despite of a relatively low surface density, the
binary fraction in this study is too high to be an intrinsic feature, if 
compared with that of other clusters. Foreground stars which cannot be excluded by
the photometric membership criteria may be the source of high value of binary fraction.
However, as mentioned in Section 3.2, it is impossible to exclude all the field interlopers
from the photometry alone, and therefore the binary fraction obtained here cannot be a
definite value.

\subsection{BACKGROUND GROUPS}
The identification of foreground and background groups of young blue stars is important in order to 
recognize the radial distribution of stars in the Galaxy. However, these studies
are limited due to the small number of early-type stars and to the difficulties in the reddening
estimate of faint stars. As shown in Section 3, three background groups except members of
NGC 2353 are found in the ($U-B$, $B-V$) diagram.
Using $E(U-B)/E(B-V) =0.72$ the reddening of each group is determined, and then the distance moduli are
estimated by shifting the reddened ZAMS to the postion of each group in the CMDs. We present the results in
Table \ref{tab4} and in Fig. \ref{fig10}.

The distance to the less reddened group is very similar to that of NGC 2353. This
group corresponds to Group A of FHR90, although there is a slight difference
in the reddening and distance obtained here. Therefore stars belonging to
Group A are the members of the CMa OB1 association. Two highly reddened groups may be related
to B type stars distributed in the outer disks in and beyond the Perseus arm as mentioned
by FHR90. Since the B-type stars in disks can exist during a couple of crossing times at almost
the same place where the stars were formed, their current location indicates that 
the star forming region in the disks and these early type stars can also be an important
tracer of the spiral arm structure of the Galaxy.

\begin{table}[t]
\begin{center}
\caption{ Reddening and distance moduli of the background groups beyond NGC 2353\label{tab4}}
\renewcommand\arraystretch{1.5}
\begin{tabular}{cccc}
\hline \hline
    & $E(B-V)$ & $V_0-M_V (d)$ & FHR90 \\
\hline
Group A & $0.18 \pm 0.02$ & $10.5$ (1.2 kpc)& A\\
Group B & $0.65 \pm 0.02$ & $12.7$ (3.5 kpc)& B\\
Group C & $0.47 \pm 0.02$ & $12.2$ (2.7 kpc)&  \\
\hline\\
\end{tabular}
\end{center}
\end{table}

\section{Conclusion}
We have performed $UBVI$ CCD photometry of NGC 2353 for the first time, and presented in the first paper of the SOS project.
We estimated a reddening of $E(B-V) = 0.10 \pm 0.02$, the cluster distance of $1.17 \pm 0.04$ kpc, 
and the cluster age of 120 million year. Both reddening and distance are in good agreement with previous studies, 
while the estimated age is slightly older than that of FHR90. From our estimate of the age of NGC 2353, we support the idea that there is no physical
connection between NGC 2353 and the CMa OB1. We derived the mass and luminosity functions of the cluster. We also estimated the binary fraction
of $48 \pm 5 \%$ and found three background groups by identifying the bright B stars along
the line of sight.

\acknowledgments{We express deep thanks to the anonymous referee for many useful comments and suggestions.
This work is supported by the Astrophysical Research Center for the Structure and Evolution
for the Cosmos (ARCSEC$\arcsec$)
at Sejong University.}

\end{document}